\documentclass{article}

\usepackage[dblblindworkshop,final]{neurips_2026}

\usepackage[utf8]{inputenc}
\usepackage[T1]{fontenc}
\usepackage{hyperref}
\usepackage{url}
\usepackage{bookmark}
\usepackage{booktabs}

\usepackage{amsmath,amssymb}
\usepackage{graphicx}
\usepackage{microtype}
\usepackage{xcolor}
\usepackage{tabularx}

\workshoptitle{Machine Learning and the Physical Sciences (ML4PS 2026)}
\title{Beyond Nested Sampling: Amortized Bayesian Model Comparison for Exoplanet Atmospheres}

\author{%
  \parbox{\linewidth}{\centering
    Nicolas~Waehner$^{1,2}$, Carolina~Cuesta-Lazaro$^{3,4}$, Daniel~Muthukrishna$^{2,5}$,
    Ana~Glidden$^{2}$, Cecilia~Garraffo$^{2}$ \\[1ex]
    {\normalfont\small
    \mbox{$^{1}$Department of Physics, Universidad de Buenos Aires},
    \mbox{$^{2}$Center for Astrophysics $|$ Harvard \& Smithsonian},
    \mbox{$^{3}$Flatiron Institute},
    \mbox{$^{4}$New York University},
    \mbox{$^{5}$Massachusetts Institute of Technology}
    }
  }
}
\newcommand{\Mzero}{{M}_0}
\newcommand{\Mone}{{M}_1}

\begin{document}

\maketitle

\begin{abstract}

The search for life on other planets hinges on a single number: the statistical significance of a biosignature gas detection. Quantifying it robustly is hard: signals may sit close to the noise level, different molecules overlap in wavelength, and the Bayesian analysis needed to account for every plausible absorber becomes intractable as the number of free parameters grows, with nested sampling, the standard tool, becoming impractical beyond roughly 30 parameters. In practice, simpler models are used instead, at the risk of spurious detections. Here, we introduce a scalable machine learning alternative to nested sampling for Bayesian model comparison in exoplanet atmosphere retrieval: a Savage-Dickey density ratio (SDDR) method based on Neural Ratio Estimation (NRE). A single network returns both the marginalized posterior of a given molecule and the Bayes factor $K$ for its presence, even when marginalizing over many other parameters. We compare it to three alternative neural estimators, Flow Matching, Evidence Networks, and SDDR with Neural Posterior Estimation (NPE), and find it to be significantly more accurate. In a controlled three-parameter scenario where the true Bayes factor is known, SDDR-NRE has the lowest error and stays calibrated up to $\ln K = 10$, while the alternatives break down at or before $\ln K \approx 5$. It then scales to a twenty-one-parameter atmosphere, where it reproduces the nested sampling posteriors and recovers $\ln K$ with an RMSE of $0.99$ nats in about 5 ms per spectrum, against a nested sampling reference whose own runs disagree by $0.61$ nats and take 2.5-11 hours per spectrum. This method is a step towards detection significances that account for every plausible absorber, and for the robust detection of life on other planets.

\end{abstract}

\section{Introduction}

Bayesian inference has become the standard framework for exoplanet atmosphere retrieval \citep{Madhusudhan_2018,Madhusudhan_2011, Line_2013}, transforming an observed spectrum $x$ into constraints on molecular abundances, atmospheric thermal structure, and cloud and haze properties, etc. Establishing the confidence with which a molecule is detected, however, has proven challenging \citep{Welbanks2026NatAs}. The current state-of-the-art approach is a leave-one-out procedure: given two models, ${M}_1$ with parameters $\phi$, and ${M}_0$ with the same parameters but excluding the molecule whose log abundance is $\theta$, the evidence for or against the additional component is summarized by the Bayes factor $K$,
\begin{equation}
  K (x)=\frac{p(x\mid\Mone)}{p(x\mid\Mzero)},\qquad
  p(x\mid M)=\int p(x\mid\phi, M)
  \pi(\phi\mid M)\,d\phi ,
  \label{eq: bayesfactor}
\end{equation}
where the evidence $p(x\mid{M})$ averages the likelihood over the full prior of model $M$. Nested sampling (NS) estimates this integral reliably but is sequential by design, so it must be
repeated for every observation and every model \citep{skilling2006,speagle2020}, at hours to days per run \citep{retrievals_take_time}. This puts repeated noise studies, end-to-end validation and retrievals carrying dozens to hundreds of molecules out of practical reach. 

Given the overlapping absorption signatures of many molecules \citep{niraula2025titan}, significance can be spuriously attributed to any one of several species absorbing in a given wavelength regime. Such degeneracies can produce misinterpretations, for example, the claimed detection of biosignature gases (DMS, DMDS) in the atmosphere of K2-18b \citep{dmsdmds}, where the inferred significance proved highly sensitive to the set of molecules included in the retrieval and dropped substantially once a broader range of plausible absorbers was considered \citep{dms_reanalysis,dms_reanalysis2}.

Avoiding such false positives requires marginalizing over every plausible absorber, which quickly pushes retrievals beyond what NS can handle. Simulation-based inference (SBI) offers a way out: it amortizes the cost of each observation via an offline training set sampled from the prior \citep{cranmer2020,dax2026introductionbayesianfrequentistsimulationbased}. In the context of exoplanet atmospheres, \cite{vasist2023} first demonstrated for simulated spectra that Neural Posterior Estimation (NPE) with normalizing flows reproduces nested-sampling posteriors in seconds, and \cite{gebhard2025} later achieved the same with Flow Matching Posterior Estimation (FMPE). NPE has since been applied to real JWST observations of brown dwarfs \citep{vasist2025}, and normalizing-flow-based SBI won the Ariel Data Challenge 2023 \citep{aubin2023}. Beyond parameter estimation, \cite{lueber2025} used amortized inference for model comparison and recovered model probabilities matching those from nested sampling. These efforts have focused on validating the inference machinery. Here, we build on past efforts to systematically assess how different SBI approaches can be used to recover molecular detection significance. With SBI, we can both accelerate calculations, saving time and computational resources, and marginalize over several dozen parameters. In this paper, we deploy and compare four approaches for estimating the H$_2$O Bayes factor in a controlled three-parameter atmospheric model, and carry the best of them to a twenty-one parameter scenario.

\section{Controlled exoplanet experiment}\label{sec: controlled_exoplanet_experiment}
We adopt the {\tt POSEIDON} radiative-transfer model and retrieval framework \citep{macdonald2023} to generate simulated transmission spectra, covering wavelengths relevant to the JWST/NIRSpec PRISM instrument at a resolution of $R = 90$, binned onto 207 channels spanning $0.53$--$5.34\,\mu$m. We inject Gaussian noise at the $50\,\mathrm{ppm}$ level, leaving the application of instrument noise to later studies so that we can focus on comparing methods. We consider two setups of increasing complexity: one with three parameters and another one with twenty-one parameters. In each case, $M_0$ is obtained by removing water. Details about implementation are in App. \ref{app: exoplanet parameters}. 

\section{Four amortized routes to one Bayes factor}
\label{sec:methods}
The four estimators all target the Bayes factor $K$ of the nested comparison defined in Eq. \ref{eq: bayesfactor}, where $\theta=\log_{10} X_{\mathrm{H_2O}}$ is the log abundance of the molecule under test and the smaller model $M_0$ is recovered by removing water. The methods differ mainly in what quantity the network must learn and the loss function to achieve that. Details about the neural networks used for each method are in App. \ref{app: neuralnetworks}.

\textbf{Flow Matching} (FM) learns a velocity field that maps a Gaussian distribution onto the distribution of spectra under $M_1$, drawing all atmospheric parameters from their priors \citep{lipman2023flowmatchinggenerativemodeling}. Following the condition-dropout strategy used in classifier-free guidance \citep{ho2022classifierfreediffusionguidance}, we mask the H$_2$O condition in 50\% of the training examples. This allows a single velocity network to learn both the H$_2$O-conditional and H$_2$O-marginalized spectral distributions. We can then obtain $p(x|M_1)$ and $p(x|\log_{10} X_{H_2O} = -12,M_1) \approx p(x|M_0)$ using the continuity equation described in \cite{lipman2023flowmatchinggenerativemodeling}, and recover the Bayes factor as the ratio of the two. Details are in App. \ref{app: flowmatching}.

\textbf{Evidence Network} (EN) uses the leaky parity-odd power (l-POP) exponential loss \citep{evidence_networks}, \mbox{$\mathcal{L}(f,m)=e^{\left[\left(\frac{1}{2}-m\right)J_\alpha(f)\right]}$} with \mbox{$J_\alpha(f)=f+f|f|^{\alpha-1}$}, model label \mbox{$m\in\{0,1\}$}, \mbox{$\alpha=2$} and \mbox{$f(x)$} the network output, whose optimum under equal model priors satisfies \mbox{$J_\alpha(f^*(x))=\ln[p(x\mid M_1)/p(x\mid M_0)]=\ln K$}. Training therefore requires only equal numbers of spectra simulated under $M_1$ and $M_0$, with parameters drawn from the priors and each spectrum labeled by its model. No evidence evaluations are needed during training; once trained, the network estimates $\ln K$ for a new spectrum in a single forward pass.

\textbf{Savage-Dickey density ratio} (SDDR) gives the Bayes factor between two nested models without computing either evidence \citep{dickey1971}. If $M_0$ is the special case of $M_1$ obtained by fixing the parameter of interest $\theta$ to a null value $\theta_0$, and the prior factorizes as $\pi(\theta, \eta) = \pi(\theta)\,\pi(\eta)$ with the nuisance prior $\pi(\eta)$ common to both models, then
\begin{equation} K = \frac{\pi(\theta_0)}{p(\theta_0 \mid x)}, \label{eq:sddr}
\end{equation}
where $p(\theta \mid x)$ is the marginal posterior of $\theta$ under $M_1$. The identity is exact, replacing a difference of two large log-evidences with a single normalized density evaluated at one point. Here $\theta = \log_{10} X_{\mathrm{H_2O}}$ and $M_0$ is the water-free atmosphere. Because the forward model returns effectively the same spectrum for every abundance below detectability, the likelihood is flat there and Eq.~\ref{eq:sddr} returns the same $K$ for any $\theta_0$ in that range.

\textbf{Neural Posterior Estimation} (NPE) approximates the posterior directly: we train a conditional normalizing flow \citep{normflows} $q_\psi(\theta \mid x)$ on pairs $(\theta, x)$ simulated from $M_1$, maximizing $\mathbb{E}[\log q_\psi(\theta \mid x)]$. The flow is one-dimensional, modeling only $\theta = \log_{10} X_{\mathrm{H_2O}}$; since the nuisance parameters are drawn from their prior in every training pair, it converges to the marginal posterior $p(\theta \mid x)$. This gives the denominator of Eq.~\ref{eq:sddr} directly, so $\ln K = \ln \pi(\theta_0) - \ln q_\psi(\theta_0 \mid x)$ in a single forward pass at negligible cost. We evaluate at $\theta_0 = -11$ to avoid hard boundary steps.

\textbf{Neural Ratio Estimation} (NRE) trains a classifier to separate joint pairs $(x, \theta) \sim p(x, \theta)$ from marginal pairs $(x, \theta) \sim p(x)\,\pi(\theta)$, learning the likelihood-to-evidence ratio $r(\theta, x) = p(\theta \mid x)/\pi(\theta) = p(x \mid \theta)/p(x)$ \citep{hermans2020, miller2021}. Balanced batches pair the simulator's output against the same spectra with $\theta$ permuted by a batch derangement, giving draws from $\pi(\theta)$ independent of $x$ at no extra simulation cost; both classes come from $M_1$. The network $h_\psi(\theta, x)$ is trained as a binary classifier: $\sigma(h_\psi)$ is the predicted probability that
a pair is a joint pair, so $h_\psi$ is the log-odds of that event. Minimizing the binary cross-entropy makes these
log-odds equal to $\log r(\theta, x)$ at the optimum \citep{hermans2020}, and at inference $h_\psi$ is used directly as $\log r$. As in NPE, the nuisance parameters are marginalized by the training itself, and the Bayes factor follows from Eq.~\ref{eq:sddr} as $\ln K = -h_\psi(\theta_0, x)$ at $\theta_0 = -11$. NRE thus estimates the log ratio itself rather than the posterior density at $\theta_0$.

\begin{figure}[h]
  \centering
  \includegraphics[width=1\linewidth]{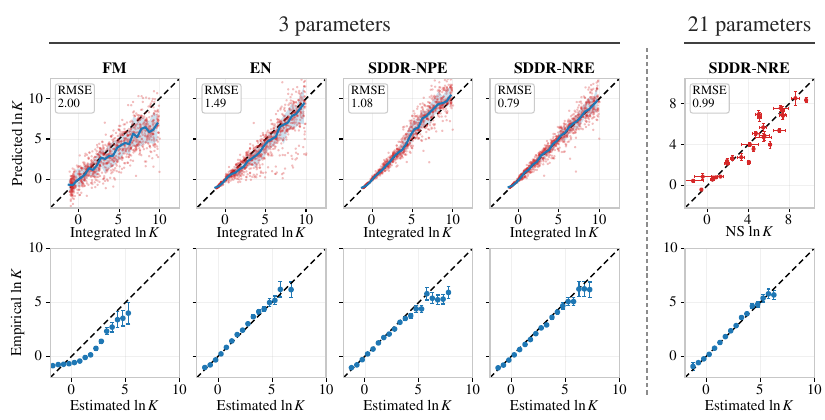}
    \caption{\textbf{Accuracy (top row) and calibration (bottom row) of the estimated Bayes factor.} Columns~1--4 show the 3-parameter problem for FM, EN, SDDR-NPE and SDDR-NRE; column~5 shows an SDDR-NRE ensemble on the 21-parameter problem. \emph{Top:} predicted $\ln K$ against the corresponding ground truth reference. In columns 1--4 red points are 990 individual spectra, the blue curve is the running median in $0.5$-nat bins of the true $\ln K$ and the shaded band the corresponding $16$th--$84$th percentile range. Column~5 plots the five-network ensemble against the mean of three independent \texttt{dynesty} runs for 28 spectra. \emph{Bottom:} empirical against estimated $\ln K$. The empirical value per $0.5$-nat bin is $\ln(n_1/n_0)$, the log ratio of $M_1$ to $M_0$ counts, from $5 \times 10^{5}$ spectra per model ($2 \times 10^{4}$ for FM). SDDR-NRE has the lowest RMSE of the four and stays calibrated in both problems.}
    \label{fig:accuracy_and_calibration_4_methods_and_20_params}
\end{figure}
\section{Results}
Not every value of $\ln K$ is equally informative, and it strongly depends on the parameters included in the model and on its priors \citep{bayes_factor_critique}. For example, spectra with water imprints can show a large $\ln K$ under the three-parameter model that is substantially reduced under the twenty-one-parameter one due to other molecules partially mimicking the water signal. Regardless of the model, spectra whose data clearly do not require water ($\ln K < 0$), and those that overwhelmingly do ($\ln K \gg 1$), are handled correctly by any reasonable estimator; what separates methods is the intermediate regime. The Jeffreys scale \citep{jeffreys1961, kass1995}, widely adopted across the community for interpreting Bayes factors, conventionally reads $\ln K \gtrsim 5$ as strong evidence for model $M_1$ over $M_0$. We decided to relax this and test our methods up to $\ln K = 10$. 

On the controlled three-parameter problem of Section~\ref{sec: controlled_exoplanet_experiment}, direct quadrature, i.e. numerical integration, provides converged reference Bayes factors so we can measure both \emph{accuracy} and \emph{calibration} (that spectra assigned $\ln K = t$ be $e^{t}$ times more likely under ${M}_1$ than under ${M}_0$). Columns 1--4 of Figure~\ref{fig:accuracy_and_calibration_4_methods_and_20_params} show that FM is uncalibrated and has high variance, while EN and SDDR-NPE are calibrated up to $\ln K \approx 5$, with SDDR-NPE being more accurate (RMSE $1.08$ against $1.49$). SDDR-NRE performs best on both diagnostics: its median follows the ground truth, it is calibrated, and its RMSE of $0.79$ is the lowest of the four. This conclusion holds using the mean absolute error instead of the RMSE. Our interpretation is that NRE outperforms NPE because of how each method learns the posterior density at $\theta_0$. NPE minimizes the KL divergence, which weights density errors by the posterior itself, so errors where the posterior is small barely affect the loss. NPE also draws its training samples from that posterior, so almost none fall near $\theta_0$. When the molecule is present, the flow's density at $\theta_0$ is therefore set by extrapolating its tail under the normalization constraint, and is systematically underestimated, which gives a positive $\ln K$ bias that grows with $\ln K$. Meanwhile, NRE is trained with contrastive pairs whose parameters are redrawn from the prior, which places samples near $\theta_0$ for every spectrum. The ratio is thus learned locally, exactly where it is read out.

\begin{figure}[h]
  \centering
  \includegraphics[width=0.8\linewidth]{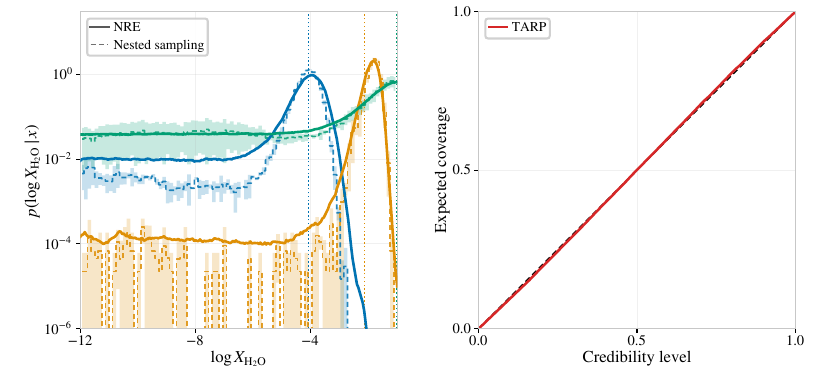}
\caption{\textbf{Marginal H$_2$O posteriors and posterior calibration for the NRE ensemble on the 21 parameter problem.} \emph{Left:} the marginal posterior of $\log_{10} X_{\mathrm{H_2O}}$ for three test spectra, with true values as vertical dotted lines. Solid curves are the ensemble of five networks and dashed curves are the binned mean of the three independent \texttt{dynesty} runs. The shaded envelope is $\pm 1$ standard deviation of those three runs. \emph{Right:} TARP coverage of the ensemble posterior.}
  \label{fig:posteriorNRE}
\end{figure}

We then tested the best-performing method, SDDR-NRE, on the 21-dimensional problem\footnote{We also tested SDDR-NPE and EN on the twenty-one parameter scenario and found the same relative performance as on the three-parameter case compared to SDDR-NRE.} and trained an ensemble of five networks of identical architecture. Reference Bayes factors are obtained three times with different \texttt{dynesty} configurations \citep{speagle2020} (details in App. \ref{app: dynesty}) to reduce the reference error. The fifth column of Figure~\ref{fig:accuracy_and_calibration_4_methods_and_20_params} therefore plots their mean, with error bars showing its standard error, and the ensemble
prediction with the spread of its members. The method loses some accuracy (RMSE = $0.99$ nats), remains calibrated, and the ensemble outperforms each network independently. SDDR-NRE also provides the marginalized molecular-abundance posterior: Figure~\ref{fig:posteriorNRE} shows that the ensemble accurately reproduces the average NS posteriors and the TARP test \citep{lemos2023tarp} shows a good calibration. The tails are difficult to compare because they are poorly sampled by NS in some cases, as shown by the orange curve. The blue curve is one of the largest deviations in our test set: the ensemble overestimates the posterior density on the low-abundance region, where the SDDR read-out is evaluated, resulting in an underestimated $\ln K$.

The ensemble of five SDDR-NRE took 5 GPU-hours to train (each network on its own V100), while the 512 million spectra used as training data took 10 GPU-hours. The three nested-sampling ground truths took around 17 h per spectrum (M0 + M1). SDDR-NRE then recovers ln K for any new spectrum in 5 ms. We expect the amortization to be even higher as dimensionality increases.

\section{Discussion and future work}

We find that SDDR-NRE is the best performing method, losing little accuracy when expanding from three to twenty-one parameters, an effect that may be partly due to scatter in the NS reference. This method could be trained for every atmospheric parameter, each time with different instrumental noises or in situations with intractable likelihoods, amortizing both the marginal posterior inference and the significance of each parameter. For significance to be assigned robustly, all plausible absorbers must be considered, a parameter regime where standard retrievals are prohibitively expensive. SDDR-NRE might be able to overcome this problem, marginalizing over all parameters during training. This is crucial for detecting degenerate atmospheric properties that could ultimately help determine whether life exists elsewhere. Future work will explore conditioning the network on a molecule index, so that a single forward pass yields the marginal posterior and the Bayes factor for every absorber, and its performance on real data.

\section{Acknowledgments}
This work was supported by the Alfred P. Sloan Foundation under grant number G-2025-79236. N.W. was supported by CONICET. We thank the members of the AstroAI group and Rocco Di Tella for useful discussions, and we thank AstroAI for providing computational resources. This work also used Bridges-2 at Pittsburgh Supercomputing Center through allocation PHY240105 from the Advanced Cyberinfrastructure Coordination Ecosystem: Services \& Support (ACCESS) program \citep{10.1145/3569951.3597559}, which is supported by U.S. National Science Foundation grants \#2138259, \#2138286, \#2138307, \#2137603, and \#2138296. This work benefited from the 2026 EXOMINTS workshop held at Ringberg Castle, Germany.

\clearpage
\bibliographystyle{plainnat}
\bibliography{references}

\clearpage
\appendix

\section{Exoplanet experiment parameters} \label{app: exoplanet parameters}

Throughout, the atmosphere is cloud-free and isothermal with an H$_2$/He bulk composition set at a helium fraction $X_{\mathrm{He}}/(X_{\mathrm{He}}+X_{\mathrm{H_2}}) = 0.17$, and the system geometry is held fixed ($R_p =
1.279\,R_{\mathrm{J}}$ at a reference pressure of $10$~bar, surface gravity $g =
10\,\mathrm{m\,s^{-2}}$, $R_\star = 1.155\,R_\odot$, $T_\star = 6071\,\mathrm{K}$).
Transmission spectra are computed on a 41-level pressure grid spanning
$10^{-6}$--$10^{2}$~bar and binned to the $207$ JWST/NIRSpec PRISM channels
between $0.53$ and $5.34\,\mu$m, with independent Gaussian noise of $50$~ppm per channel. We reimplemented the forward model in JAX \citep{jax2018github} with the opacities pre-interpolated onto a fixed temperature--pressure--wavelength grid, so that generating a spectrum reduces to a jitted lookup and matrix contraction executed on the GPU. This costs roughly $0.1\,$ms on an H100 per spectrum against $\sim\!17\,$ms on a CPU core, resulting in a major speed-up.

In the \emph{three-parameter} case the larger model ${M}_1$ has the base-10 logarithms of the volume mixing ratios (VMRs), $\log_{10} X_{\mathrm{H_2O}}$ and $\log_{10} X_{\mathrm{CO_2}}$, and the atmospheric temperature $T$ in kelvin; the smaller model deletes water and retains the latter two. The \emph{twenty-one-parameter} case extends this to twenty absorbers --- H$_2$O, CO$_2$, CO, Na, K, SO$_2$, H$_2$S, CH$_4$, NH$_3$, HCN, C$_2$H$_2$, PH$_3$, SiO, TiO, VO, FeH, CrH, AlO, MgH and N$_2$O --- each with a free log-abundance, plus temperature, so that ${M}_1$ has 21 parameters and ${M}_0$, which again deletes water, has 20. Log-abundances have uniform priors on $[-12,-1]$ and  $T \sim \mathcal{U}(500, 2000)\,\mathrm{K}$.

\section{Estimator architectures}\label{app: neuralnetworks}

All methods used the same 207-bin spectra and fixed per-bin standardization. Table~\ref{tab:network_architectures} summarizes the architectures used in the three-parameter comparison. Each model was trained with Adam for $10^6$ steps and final weights were retained.

\begin{table*}[h]
    \centering
    \small
    \setlength{\tabcolsep}{3pt}
    \begin{tabularx}{\textwidth}{@{}lXlrrr@{}}
      \toprule
      Method & Architecture & Loss & Parameters & Batch & Initial LR \\
      \midrule
      EN  & 6-block CNN (64--128 channels) + MLP (256)
          & l-POP Exp. ($\alpha=2$) & 387,713 & $2\times256$ & $10^{-4}$ \\

      FM  & 10-block FiLM CNN (320 channels)
          & Conditional OT-FM & 10,666,433 & 256 & $2\times10^{-4}$ \\

      NPE & 6-block CNN (context 64) + NSF (3 transforms, $2\times128$ hidden units, 16 bins)
          & NLL & 496,589 & 512 & $3\times10^{-4}$ \\

      NRE & 6-block CNN (context 64) + MLP ($256\times2$) with 16 Fourier features
          & BCE & 495,041 & $512^\dagger$ & $3\times10^{-4}$ \\
      \bottomrule
    \end{tabularx}
    \caption{Architectures and training settings used for the four estimators.
    Parameter counts include all trainable components.
    $^\dagger$The 512 simulated spectra produce 1024 joint and marginal classifier pairs.}
    \label{tab:network_architectures}
\end{table*}

For EN, NPE, and NRE, the learning rate was multiplied by $0.9988$ every 1000 steps. FM used a \texttt{ReduceLROnPlateau} schedule with a reduction factor of $0.5$. The NPE and NRE gradients were clipped to a norm of 10.

The 21-dimensional ensemble results in the fifth column of Figure~\ref{fig:accuracy_and_calibration_4_methods_and_20_params} use the same NRE architecture listed above for every network. Each network was retrained from scratch for $10^6$ steps. We normalized the data with statistics from 20,000 simulations drawn from the 21-dimensional prior.

\section{Flow Matching}\label{app: flowmatching}
We trained a single masked conditional flow using spectra simulated exclusively under $M_1$. For each simulation, all atmospheric parameters were independently drawn from their priors. The flow learns a velocity field $v_{\psi}(x_t,t\mid c, m)$,
where $c=\log_{10} X_{\mathrm{H_2O}}$ and $m$ indicates whether this value is masked. In half of the training examples, the true H$_2$O abundance is supplied to the network, allowing it to learn $p(x\mid \log_{10} X_{\mathrm{H_2O}},M_1)$,
with the remaining atmospheric parameters marginalized over their priors. In the other half, the H$_2$O abundance is masked, so the flow learns the distribution obtained after marginalizing over all the parameters of $M_1$, namely $p(x\mid M_1)$. The learned probability path satisfies the continuity equation \citep{lipman2023flowmatchinggenerativemodeling}. Starting from a standard Gaussian density $p_0$, we evaluate the density of an observed spectrum by integrating the learned flow backwards from the spectrum to its corresponding Gaussian latent variable:
  \[
  \log p_{\psi}(x\mid c,m)
  =
  \log p_0(x_0)
  -
  \int_0^1
  \nabla_x\cdot v_{\psi}(x_t,t\mid c,m)\,dt.
  \]
We compute this integral using the adaptive ODE solver Dormand--Prince RK45 \citep{dormand_integrator}. The two model evidences are then obtained from the two operating modes of the same network,
  \[
  \log Z_1=\log p_{\psi}(x\mid m=1),
  \qquad
  \log Z_0\simeq\log p_{\psi}
  \left(x\mid c=\log_{10} X_{\mathrm{H_2O}}=-12,m=0\right).
  \]
where a log-abundance of $-12$ produces a negligible H$_2$O imprint and therefore approximates the genuinely H$_2$O-free model $M_0$. Finally, the Bayes factor is
  \[
  \log K=\log Z_1-\log Z_0.
  \]

\section{Nested Sampling reference}\label{app: dynesty}

At 21 parameters no exact reference is available, so Bayes factors are computed with \texttt{dynesty} \citep{speagle2020} in three configurations of increasing cost: 2000 live points with slice sampling at its default of $3 + n_{\mathrm{dim}}$ steps per likelihood replacement (24 for ${M}_1$), 2000 live points with 48 steps, and 4000 live points with 48 steps. Averaged over the 28 test spectra, one spectrum costs 16M, 31M and 62M likelihood evaluations, approximately 2, 5 and 11 hours on a V100; around 530 GPU-hours in total. The three runs disagree with one another by 0.61 nats rms, so we treat no single run as ground truth and therefore use the mean of the three. In comparison, the SDDR-NRE method takes 1 ms per network per spectrum.

\end{document}